\documentclass[
 reprint,
 amsmath,amssymb,
 aps,
superscriptaddress ]{revtex4-2}

\usepackage{graphicx}
\usepackage{dcolumn}
\usepackage{bm}
\usepackage{physics}
\usepackage[colorlinks=true, linkcolor = blue, hyperfootnotes=false, citecolor = blue]{hyperref}
\usepackage{color,soul}
\begin{document}

\preprint{APS/123-QED}


\title{Smart patterning for topological pumping of elastic surface waves}

\author{Shaoyun Wang}
\thanks{These three authors contributed equally}
\affiliation{Department of Mechanical and Aerospace Engineering, University of Missouri, Columbia, MO 65211, USA}
\author{Zhou Hu}
\thanks{These three authors contributed equally}
\affiliation{School of Aerospace Engineering, Beijing Institute of Technology, Beijing 100081, China}
\author{Qian Wu}
\thanks{These three authors contributed equally}
\affiliation{Department of Mechanical and Aerospace Engineering, University of Missouri, Columbia, MO 65211, USA}
\author{Hui Chen}
\affiliation{Piezoelectric Device Laboratory, School of Mechanical Engineering and Mechanics, \\
Ningbo University, Ningbo 315211, China}
\author{Emil Prodan}
\affiliation{Department of Physics, Yeshiva University, New York, NY 10016, USA}
\author{Rui Zhu}
\email[Corresponding author: ]{ruizhu@bit.edu.cn}
\affiliation{School of Aerospace Engineering, Beijing Institute of Technology, Beijing 100081, China}
\author{Guoliang Huang}
\email[Corresponding author: ]{huangg@missouri.edu}
\affiliation{Department of Mechanical and Aerospace Engineering, University of Missouri, Columbia, MO 65211, USA}

\date{\today}

\begin{abstract}
Topological pumping supplies a robust mechanism to steer waves across a sample without being affected by disorders and defects. For the first time, we demonstrate the pumping of elastic surface waves, achieved by a smart patterning of a surface that creates a synthetic dimension, which is explored by the wave as it is launched perpendicularly to the steering direction. Specifically, we design and fabricate an elastic medium decorated with arrays of pillar-type resonators whose eigenmodes are locate below the sound cone, together with coupling bridges edged according to a specific algorithm. We establish a connection between the collective dynamics of the pillars and that of electrons in a magnetic field by deriving an accurate tight-binding model and developing a WKB-type analysis suitable for such discrete aperiodic systems with spatially slow-varying couplings. This enable us to predict topological pumping pattern, which is numerically and experimentally demonstrated by steering waves from one edge of the system to the other. Finally, the immune character of the topologically pumped surface waves against disorder and defects is evidenced. The principle of surface patterning together with the WKB-analysis could provide a powerful new platform for surface wave control and exploration of topological matter in higher dimensions.


\end{abstract}

\maketitle
\section{Introduction}

\noindent Topological matter is a rapidly growing field in which topological concepts are exploited to discover and classify new phases of matter \cite{hasan2010colloquium,qi2011topological,chiu2016classification,rachel2018interacting}. In this context, a hallmark achievement was the discovery of the integer quantum Hall effect [citation]. In the past decade, topological phases analogous to quantum Hall insulators have been engineered across a wide range of time-modulated platforms, including electronics \cite{lindner2011floquet,xu2014observation,yoshimi2015quantum}, photonics \cite{haldane2008possible, wang2009observation,rechtsman2013photonic,hafezi2013imaging,lu2014topological,ozawa2019topological}, acoustics \cite{fleury2016floquet,yang2015topological,mousavi2015topologically,souslov2017topological}, and mechanics \cite{wang2015topological,nash2015topological,nassar2018quantization,chen2019mechanical,Rosa2019,Riva2020}. The existence of the conventional gapless edge states and surface states is guaranteed by the bulk-boundary correspondence. These time-dependent systems can provide outstanding opportunities not possible with passive materials, enabled by the high controllability and flexibility of these platforms. However, a physical realization of a dynamically controlled topological pumping that produces topological transport is extremely challenging because external or active physical fields are typically needed \cite{grinberg2020robust}.

To overcome the challenges associated with time-modulated system, rendering synthetic dimensions via space modulations was recently suggested because it does not require any active materials or other external mechanisms to break the time-reversal symmetry [19, 20]. The phases of the space-modulations can be used as adiabatic parameters that augment the physical space [22]. It is intriguing to see these phases as additional global degrees of freedom, usually called phasons, living on a torus. The central idea of synthetic dimensions is to exploit and harness such degrees of freedom with atoms, photons or phonons to mimic the dynamic motion along extra spatial directions. The key advantage of synthetic dimensions is that pumping parameters can be engineered very naturally in the strength of the couplings along the extra dimension. Synthetic dimensions have led to new discoveries of the 2D and 4D quantum Hall systems in ultracold atomic gases \cite{price2015four, lohse2016thouless, lohse2018exploring}, photonics \cite{Kraus2012, Kraus2013,verbin2015topological, zilberberg2018photonic} and acoustics and mechanics \cite{chen2021creating,Long2019,chen2021landau} due to their flexibility. Rendering of the synthetic space is growing into one of most appealing approach to control and steer topological wave transport in different systems.

Surface elastic waves are a class of polarized waves that propagate on the surface of a semi-infinite elastic medium. They are confined within a superficial region whose thickness is comparable with their wavelength \cite{rayleigh1885waves}. Manipulating surface waves has been of considerable interest with widespread applications in earthquake mitigation, nondestructive evaluation, wave filtering and sensing \cite{campbell2012surface,morgan2010surface,hashimoto2000surface}. Based on the Bragg scattering and local resonance mechanisms, manipulation and control of surface waves has been recently investigated in the phononic and metamaterial community for various applications such as exotic wave transmission and reflection, wave focusing and cloaking \cite{jin2017pillar,packo2019inverse,wu2021non}. Among existing approaches, the metamaterial with pillar-type resonators is regarded as one of the most promising microstructure designs because of their simple structure and process-friendly fabrication. However, it is not trivial to apply pillar-type metamaterials for the topological surface wave transport. Indeed, there is of fundamental and practical significance to physically realize space-modulated pillar-type metamaterials for topological surface wave transport along desired orbits \cite{wang2022extended,zhang2021topological}.

In this study, we present theoretical, numerical and experimental investigations of Rayleigh wave topological pumping by leveraging a pillar-based platform with space modulations. The proposed structures can be described as aperiodic mechanical wave-channels carrying different phason values that are stacked and coupled with each other. By slowly varying the phason along the stacking direction, we demonstrate here that, with such an approach, we can explore any continuous orbit inside the phason space, and even control the speed along the path to shape the surface pumped pattern. As a result, we can render these abstract trajectories, occurring in the synthetic dimensions, on the physical dimension along the stackings. In turn, this enables us to control the propagation of the surface waves in space as well as the temporal phases of the signals.

With the control over the phason, we experimentally demonstrate edge-to-edge topological wave pattern on the space-modulated mechanical metasurface, which is robust against random fluctuations in the couplings. The analytical study of pumping process under adiabatic condition is formulated by using the WKB approximation and the modulation functions of parameters with nontrivial topological phase is also analytically obtained. Based on that, we further explore various ways in which we can control these pumping processes and validate topological mode steerings in time-domain simulation. It is believed that our work breaks ground for engineering applications, where the couplings in a space-modulated mechanical metasurface can be programmed for selective and robust point-to-point transport of surface wave signals.

\begin{figure*}
  \centering
  \includegraphics[width=\linewidth]{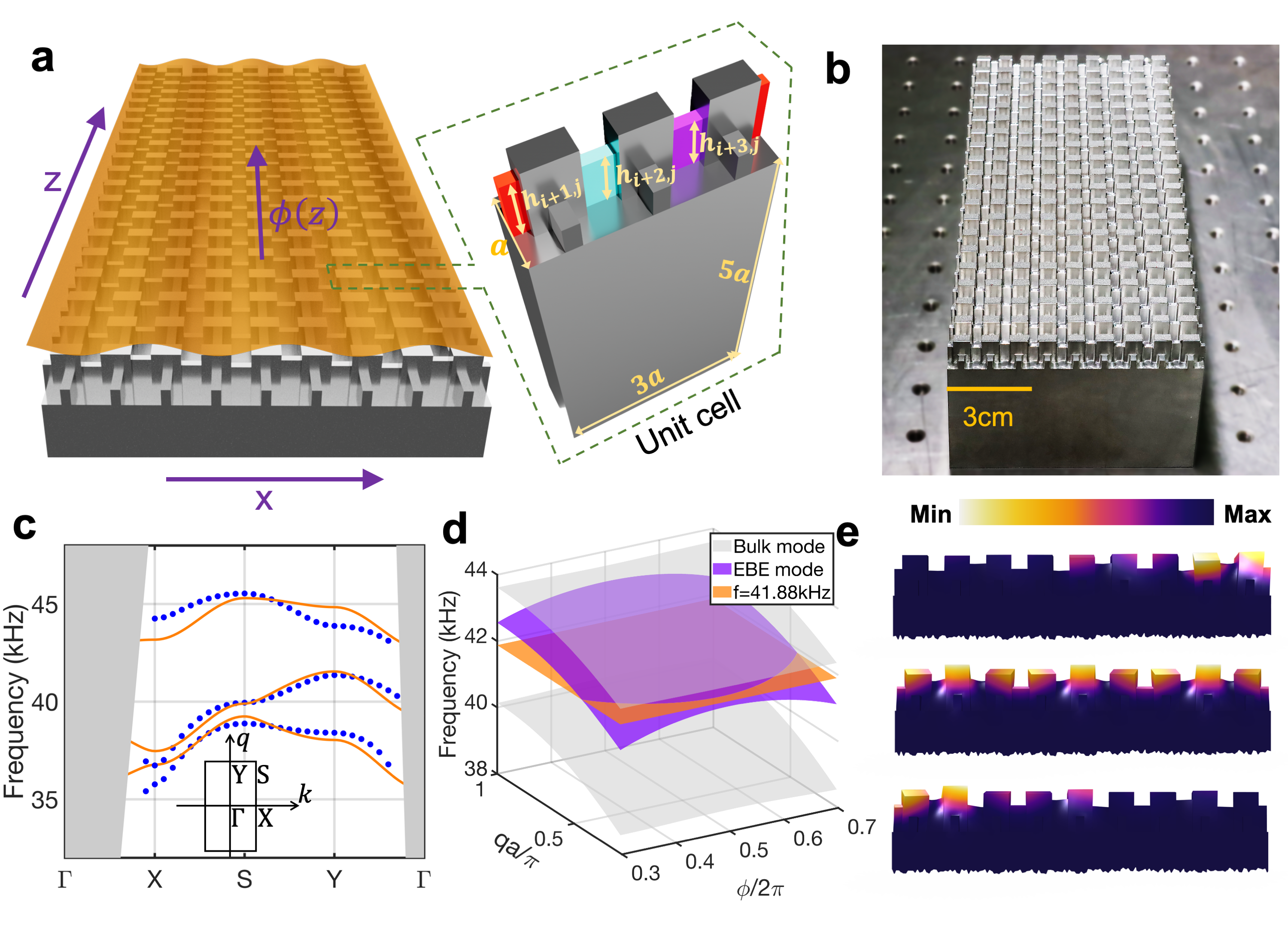}
  \caption{\textbf{Design principle and dispersion analysis}. \textbf{a} Schematic illustration of the topological surface wave transport system. Each row in $x$ corresponds to a supercell that includes three unit cells (inset). \textbf{b} Photograph of the experimental sample fabricated out of aluminum by a milling machine. \textbf{c} Numerically obtained dispersion curves (blue dots) for the unit cell with $\phi_j=\pi$. The orange curves represent the dispersion curves of the discrete mass-spring model obtained by numerical fitting. The gray regions are filled with bulk modes. Their interfaces with the surface wave region define the sound cone. \textbf{d} Dispersion diagram for the supercell terminated by free boundary conditions in the $x$-direction and Floquet boundary conditions in the $z$-direction.  The edge-bulk-edge (EBE) mode is represented by the magenta surface, whereas the bulk bands are indicated by gray surfaces. The orange cut plane corresponds to the excitation frequency $f_c=41.88\text{Hz}$. The interaction curve between excitation frequency plane and EBE surface gives the instantaneous wave number $q(z)$ on which the circle is right edge mode, the triangle is the bulk mode and the square is the left edge mode. \textbf{e} The top, middle and bottom panels are the corresponding eigenmodes of supercell at $\phi = 0.6\pi, \, \pi, \, \text{and} \, 1.4\pi$ with $q=\pi/a$ in \textbf{d}.}
  \label{fig1}
\end{figure*}

\section{Results}

\noindent \textbf{Physical rendering of synthetic spaces.} We start by explaining the principles of physical rendering of synthetic spaces in the context of surface wave transport. Figure~\ref{fig1}a, b show our surface wave platform featuring a planar array of elastic pillar-type resonators coupled horizontally and vertically through thin plates (see \textcolor{blue}{Methods sample preparation} for fabrication details). Each resonator is assigned an address $(i,j) \in \mathbb Z^2$ in the $x$-$z$ plane. The heights of the connecting plates in the $x$-direction are modulated according to the protocol $h_{ij}=h_0[1+\Delta_0 \cos(2\pi i/3+\phi_j)]$, while the geometry of the connecting thin plate along $z$ direction is uniform across the sample. Any such modulation has a phase that can take any value in the abstract interval $[0,2\pi]$, representing here the synthetic space. In a time-modulated setting, one will dynamically drive the phase $\phi$ by rapid re-configurations of the systems \cite{ChengPRL2020}. Instead, by setting the phason value of the $j$-th row as $\phi_j (z) = \phi_s+(\phi_f-\phi_s)\frac{j}{N}$ with $N$ being the total number of rows, we effectively render the synthetic space along the $z$-axis. The parameters will be fixed as $\phi_s=0.6\pi$ and $\phi_f = 1.4\pi$.

As shown in Fig. \ref{fig1}a, each $x$-directional row displays a unit cell containing three pillars. The dispersion curves of the unit cell obtained with COMSOL Multiphysics is shown in Fig.~\ref{fig1}c. The computation was carried out by imposing Floquet boundary conditions in both $x$- and $z$-directions. Since the modulation amplitude $\Delta_0$ is small, $\phi_j$ is irrelevant for the dispersion curves and can be assumed as 0. Below the sound cone (white region), one can see three surface wave branches, whose eigenmodes are localized on the surface and decay quickly into the bulk; see also \textcolor{blue}{Supplementary Information} for detailed illustrations of the corresponding mode shapes. The region above the sound cone, shown in gray, is referred to as the "bulk modes" region.

\begin{figure*}
  \centering
  \includegraphics[width=\linewidth]{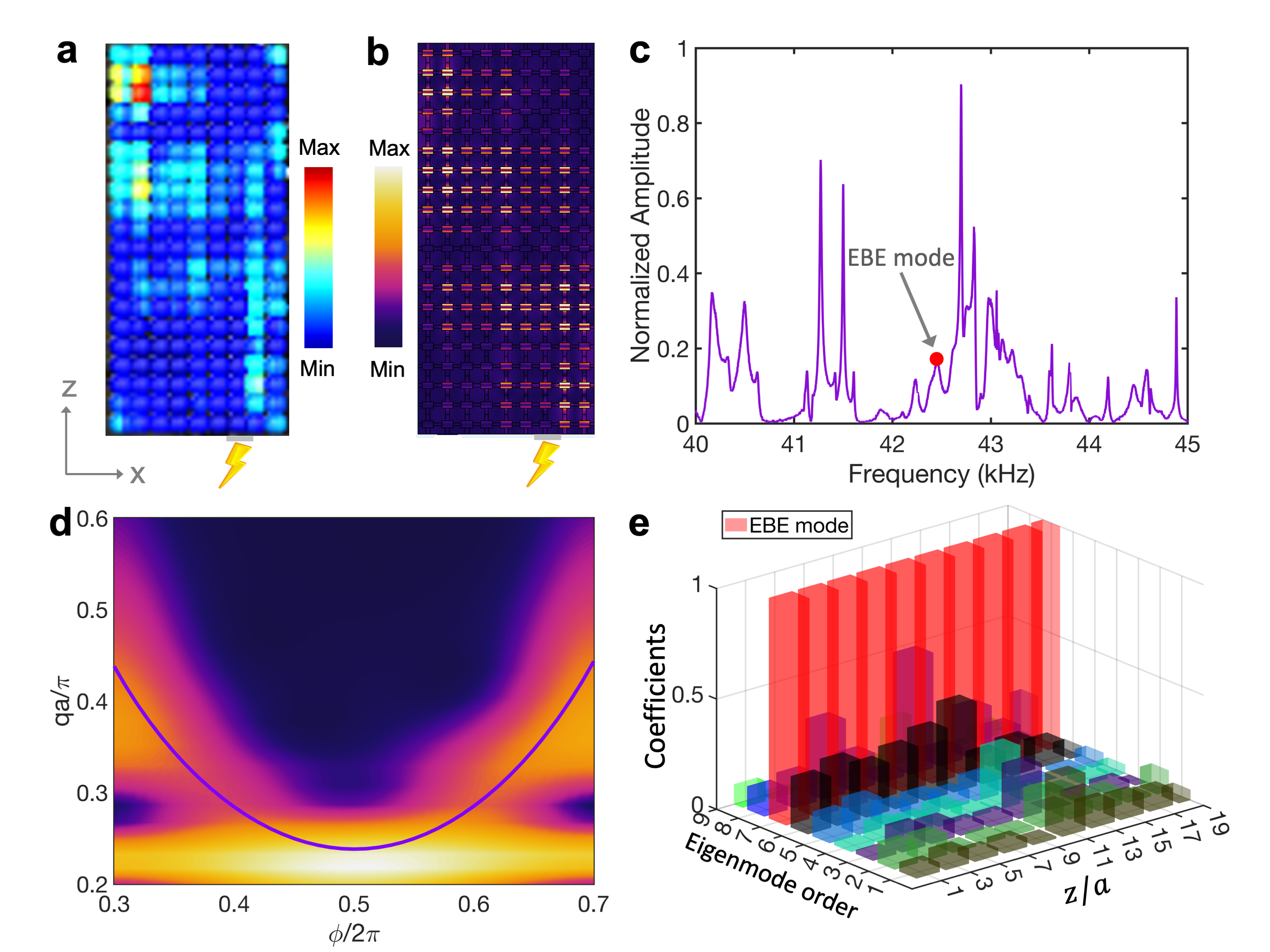}
  \caption{\textbf{Topological surface pumping on the elastic surface with space modulated pillars}. \textbf{a,b} Experimental \textbf{a} and numerical \textbf{b} modal profile of out-of-plane displacement field at the frequencies $f=$42.45 \ \text{kHz} and $f=$41.88 \ \text{kHz}, respectively, by piezoelectric patch (gray part in the bottom of the cuboid) excitation. \textbf{c} Frequency spectrum of the resonator with indices $i=2 \ \text{and} \ j=19$ from experimental measurement. The resonance peak noted by red dot is the EBE mode. \textbf{d} The wavelet transform of the eigenmode from numerical simulation along synthetic dimension. The purple curve is the interaction curve from Fig. \ref{fig1}d. \textbf{e} The mode decomposition of the displacement field in the right panel of \textbf{a} of each chain for different $z$.}
  \label{fig2}
\end{figure*}


To facilitate the physical interpretation of the surface wave pumping and illuminate the function of the phason, we develop a discrete mass-spring model for the surface wave eigenmodes using the mode-coupling theory. This model takes the form of the following difference equation for the amplitudes $\psi_{i,j}$ of the local resonances carried by the individual pilars (see \textcolor{blue}{Supplementary Information} for derivation details)
\begin{equation}\label{gover_eq}
  \begin{aligned}
   & \kappa^0 \psi_{i,j} + \kappa^h_{i-1,j}\psi_{i-1,j} + \kappa^h_{i,j} \psi_{i+1,j} \\
   & \quad + \kappa^v [\psi_{i,j-1} -2\psi_{i,j}+\psi_{i,j+1} ] = -M \omega^2 \psi_{i,j}.
  \end{aligned}
\end{equation}
Here, $M$, $\kappa^0$, $\kappa^v$ and $\kappa^h$ are interpreted as the effective mass and grounded, vertical and horizontal spring stiffnesses of the model, respectively. The values of these effective parameters are determined by fitting the dispersion curves of the continuous model (blue dots in Fig.~\ref{fig1}c). Specifically, we obtain $M=1~\text{kg}$, $\kappa^0=49.6~\text{GN/m}$,  $\kappa^v=1.9~\text{GN/m}$, and $\kappa^h_{i,j} = \kappa^h_0\left[1+\Delta\cos(2i\pi/3+\phi_j)\right]$, where the modulation coefficients read $\kappa^h_0=5.5 ~\text{GN/m}$ and $\Delta=0.67$. As shown in Fig.~\ref{fig1}c, the continuous and discrete dispersion curves exhibit satisfactory agreement, thereby demonstrating the reliability of the discrete model.

\noindent \textbf{WKB-type analysis.} By replacing the index $j$ with the coordinate $z=ja$, we rewrite $\psi_{i,j}=\psi_i(z)$ and $\phi_j=\phi(z)$, as well as
\begin{equation}\label{Eq:KappaH}
\kappa^h_{i,j}=\kappa^h_{i}(z)= h_0[1+\Delta_0 \cos(2\pi i/3+\phi(z))].   
\end{equation}
We also introduce the second-order central difference operator 
\begin{equation}\label{delta_operator}
  \delta^2 f(z)= \frac{f\left(z+a\right)-2f(z)+f\left(z-a\right)}{a^2}.
\end{equation}
In addition, a vector $\boldsymbol{\psi}(z) = [\psi_{0}(z),\psi_{1}(z),...,\psi_{3M}(z)]^\text{T}$ is defined including all the mode coefficients. By doing so, the dispersion equation~\eqref{gover_eq} can be written very compactly as 
\begin{equation}\label{WKB equation}
  a^2 \kappa^v \delta^2 \boldsymbol{\psi}(z) + [\textbf{K}(z) + \omega^2]\boldsymbol{\psi}(z)= 0,
\end{equation}
in which $\textbf{K}(z)$ is the matrix with the entries 
\begin{equation}\label{stiff_matrix}
  \text{K}_{ik}(z) = \kappa^0 \delta_{ik} + \kappa^h_{i}(z)\delta_{i,k+1} + \kappa^h_{i}(z)\delta_{i+1,k},
\end{equation}
where $\delta_{ik}$ is the Kronecker delta. Equation~\eqref{WKB equation} is very close in spirit with the Schroedinger equation appearing in the setting of WKB approximation theory \citep{holmes2012introduction,nassar2018quantization}. The difference is that, instead of dealing with a potential, we are dealing with the non-diagonal matrix \textbf{K}(z) which, nevertheless, is slowly varying with $z$. In this regime, the following WKB-type expansion is justified
\begin{equation}
  \boldsymbol{\psi}(z) = e^{i \theta(z) /a } [\boldsymbol{\psi}^{(0)}(z)(z) + a\boldsymbol{\psi}^{(1)}(z)(z)+ \cdots],
\end{equation}
and, by keeping track of the powers of $a$, we can derive the exact equations satisfied by each $\boldsymbol{\psi}^{(\alpha)}$. In particular, we find for the leading term that this equation is (see \textcolor{blue}{Supplementary Information})
\begin{equation} \label{leading_eq}
 \big(\boldsymbol{K}(z)+\omega^2\big) \boldsymbol{\psi}^{(0)}(z) = 4 \sin^2\left(\frac{\delta \theta(z)}{2}\right) \boldsymbol{\psi}^{(0)}(z),
\end{equation}
where $\delta \theta(z)= [\theta(z+a/2)-\theta(z-a/2)]/a$. This equation has solutions of the form
\begin{equation}\label{WKB solution}
    \boldsymbol{\psi}_n(z)= A_n(z) e^{\text{i} \sum_{\xi=0}^{\xi=z} q_{n}(\xi)} \boldsymbol{\varphi}_n (z) + o(a),
\end{equation}
where $\boldsymbol{\varphi}_n (z)$ is the $n$-th eigenmode of the $\textbf{K}(z)$ matrix
\begin{equation}\label{instant_eig}
  \textbf{K}(z) \boldsymbol{\varphi}_n(z) = -\mu_n(z) \boldsymbol{\varphi}_n(z),
\end{equation}
at row $z$ and $q_n(z)$ satisfies the equation
\begin{equation} \label{frequency cutting}
  4\kappa^v \sin ^{2} \frac{q_{n}(z)}{2}+\mu_{n}(z)=\omega^{2}.
\end{equation}
As in the standard WKB-theory~\citep{holmes2012introduction}, an analysis at the order-one level of the asymptotic expansion enables us to pinpoint the $z$-dependence amplitude $A_n(z)$ (see \textcolor{blue}{Supplementary Information}), and to finally present the complete set of solutions for the dispersion equation~\eqref{WKB equation} 
\begin{equation}\label{WKB solution}
    \boldsymbol{\psi}_n(z)= \frac{c_n}{\sqrt{\omega^2-\mu_n(z)}} e^{\text{i} \sum_{\xi=0}^{\xi=z} q_{n}(\xi)} \boldsymbol{\varphi}_n (z) + o(a),
\end{equation}
We recall that the derivation of these solutions relies only the adiabatic evolution of the phason with $z$ and no considerations of long-wavelengths or paraxial approximation were made. Thus, our results cover the short-wavelength and nonparaxial regions. Lastly, since our samples are finite, we need to impose free boundary conditions on the top and bottom boundaries in the $z$-direction. In this case, the mode shape of the $n$th eigenmode is in the form of
\begin{equation} \label{general_sol}
  \boldsymbol{\psi}_n(z) =  \frac{c_n\sin Q_n(z) + d_n\cos Q_n(z) }{\sqrt{\omega^2 - \mu_n(z)}} \boldsymbol{\varphi}_n (z),
\end{equation}
where $c_n$ and $d_n$ are coefficients of superposition and $ Q_n(z) = \sum_{\xi=0}^{\xi=z} q_n(\xi)$ is the dynamical phase produced by our derivation. 


\begin{figure}
  \centering
  \includegraphics[width=\linewidth]{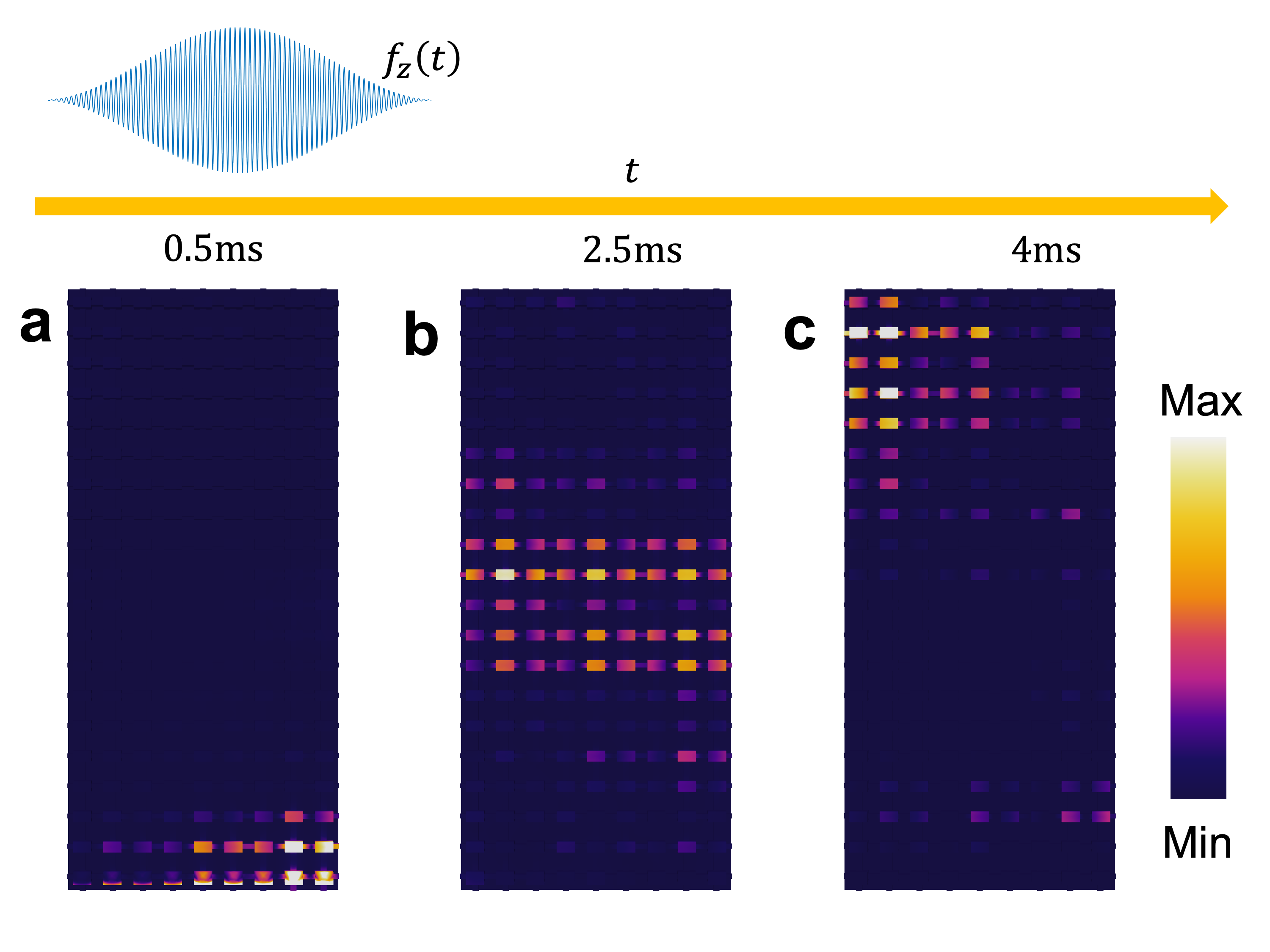}
  \caption{\textbf{Time response of the topological surface wave transport}. \textbf{a}-\textbf{c}. Magnitudes of displacement field at $0.5~\text{ms}$, $2.5~\text{ms}$ and $4~\text{ms}$, respectively. A 50-cycle tone burst excitation centered at 41.88 kHz is applied on the bottom supercell}
  \label{transient}
\end{figure}

\noindent \textbf{Topological pumping.} The complete set of solutions~\eqref{WKB solution} indicates that, when the meta-surface is excited at pulsation $\omega$ with a source placed at position $z=0$, it will resonate very strongly with the mode that has its resonant frequency $\mu_n(z=0)$ close to $\omega^2$. Thus, we have a mechanism to selectively load a specific mode out of a fairly rich set of resonant modes. Furthermore, Eq.~\eqref{WKB solution} indicates that, with such a source turned on, upon the inspection of row $z$, we will see the eigenmode $\mu_n(z)$ of the one-dimensional tight-binding operator $\textbf{K}(z)$ (up to a multiplicative factor). Since $\textbf{K}(z)$ depends only on the phason value $\phi(z)$, {\it i.e.} $\textbf{K}(z)=\textbf{K}_{\phi(z)}$, one can now see explicitly how the dependence of the spectral properties of $\textbf{K}_\phi$ on the phason has been rendered along the $z$-coordinate, for us to experience, measure and use its resonant modes in future applications. Furthermore, by design, the phason is being pumped from $\phi_i$ to $\phi_f$ as the structure is examined from bottom ($z=0$) to the top ($z = N a$).

We now turn our attention to $\textbf{K}_\phi$ defined by the equations~\eqref{Eq:KappaH} and \eqref{stiff_matrix}, and process it as
\begin{equation}\label{tstiff_matrix}
  \text{K}_{ik}(\phi) = \kappa^h_{i}(\phi)[\delta_{i,k+1} + \delta_{i+1,k}+\kappa^0 \tilde \kappa^h_{i}(\phi)\delta_{ik}] ,
\end{equation}
where
\begin{equation}
    \tilde \kappa^h_{i} = 1/\kappa^h_{i} \approx 1/h_0[1-\Delta_0 \cos(2\pi i/3+\phi(z))].
\end{equation}
Except for the multiplicative factor in front, Eq.~\eqref{tstiff_matrix} coincides with the Bloch decomposition along the $z$-direction of the Hamiltonian of electrons hopping on a two dimensional lattice in the presence of a uniform magnetic field, if the latter is rendered in the Landau gauge and the phason $\phi$ is identified with the $k_z$ quasi-momentum. The algorithm~\eqref{Eq:KappaH} used for the height of the coupling bridges sets the value of the virtual magnetic field to $\frac{1}{3}$-unit of flux per resonator. This connection enables one to effortlessly understand the topological character of the dynamics and the ensuing bulk-boundary correspondence. Specifically, the three surface dispersion bands seen in Fig.~\ref{fig1}c are the three spectral bands seen in the Hofstadter butterfly at $\frac{1}{3}$-flux \cite{NiCommPhys2019} and, as such, the spectral gaps between these bands carry Chern numbers $\pm 1$. This implies that the $x$-terminated sample will display topological edge modes which disperse with the variation of the phason. More formal analysis is supplied in the \textcolor{blue}{Supplementary Information}.

The spectrum of an entire row of resonators $q$-twisted Floquet boundary conditions imposed in the $z$-direction is reported in figure \ref{fig1}d as function of $\phi$ and $q$ and the topological edge modes can be seen as the sheet colored in purple. Taking a slice at a fixed $q$ reveals precisely one chiral edge band per edge and the slopes of these bands are consistent with the values of the Chern numbers (see \textcolor{blue}{Supplementary Information}). Furthermore, examination of the eigenfunctions leads to the observation of right edge, bulk and left edge modes for $\phi = 0.6\pi, \, \pi, \, \text{and} \, 1.4\pi$ in the top, middle and bottom panels of Fig.~\ref{fig1}e, respectively.

\noindent \textbf{Demonstration of topological surface wave transport.} We now focus on the demonstration of the topological surface wave transport. Experiments are first conducted on the system shown in Fig \ref{fig1} (see \textcolor{blue}{Methods experimental protocol}). Figure \ref{fig2}a displays the experimentally measured EBE field profile at 42.45 kHz and the harmonic numerical simulation at 41.88 kHz. In both cases, excitation is applied using piezoelectric actuators on the bottom supercell. We found the EBE mode experimentally by examining the mode shapes of all measured resonance peaks in the spectrum shown in Figure \ref{fig2}b. Vertical oscillation of the field profile is also observed featuring modal nodes and antinodes, owing to the $z$-directional dynamical phase. The experimental and numerical results provide satisfactory agreement. To quantitatively compare the retrieved mode profile in Fig. \ref{fig2}a with the analytical solution Eq. (\ref{general_sol}), we apply wavelet transform and mode decomposition on the numerical mode profile. In detail, we first divided the cuboid into 9 columns. Then, the wavelet transform technique is applied to the wave component of each column to determine the corresponding coefficients (see \textcolor{blue}{Supplementary Information}). Last, we take average on the absolute values of these coefficients. The outcome after linear interpolation is illustrated as a heat map in Fig. \ref{fig2}d. As a reference, a purple curve is given to provide the $k_z$-$\phi$ relation at 41.88 kHz on the cut plane of the dispersion diagram (Fig \ref{fig1}d). Satisfactory agreement is found between the eigenmode analysis of the finite lattice and the dispersion diagram. Next, we adopt mode decomposition on each of the 20 supercells along the $z$-direction to determine the relative strengths of all modes. Figure \ref{fig2}e illustrates the corresponding modal coefficients which are normalized with the maximum of coefficient at respective values of $z$. The bases for mode decomposition are from the corresponding mass-spring model whose parameters are extracted from Fig. \ref{fig1}c. Since only 20 supercells are involved in the synthetic dimension, the stiffness matrix does not evolve strictly adiabatically. As a result, other bulk modes always coexist. However, the EBE mode, labeled as the 7th mode in Fig. \ref{fig2}e, is always dominant in terms of modal coefficients of all the supercells, meaning that length of the synthetic dimension is sufficiently long to approach adiabatic limit. The consistence of the results from wavelet transform and mode decomposition analysis validates the correctness of the WKB solution. 

\begin{figure}
  \centering
  \includegraphics[width=\linewidth]{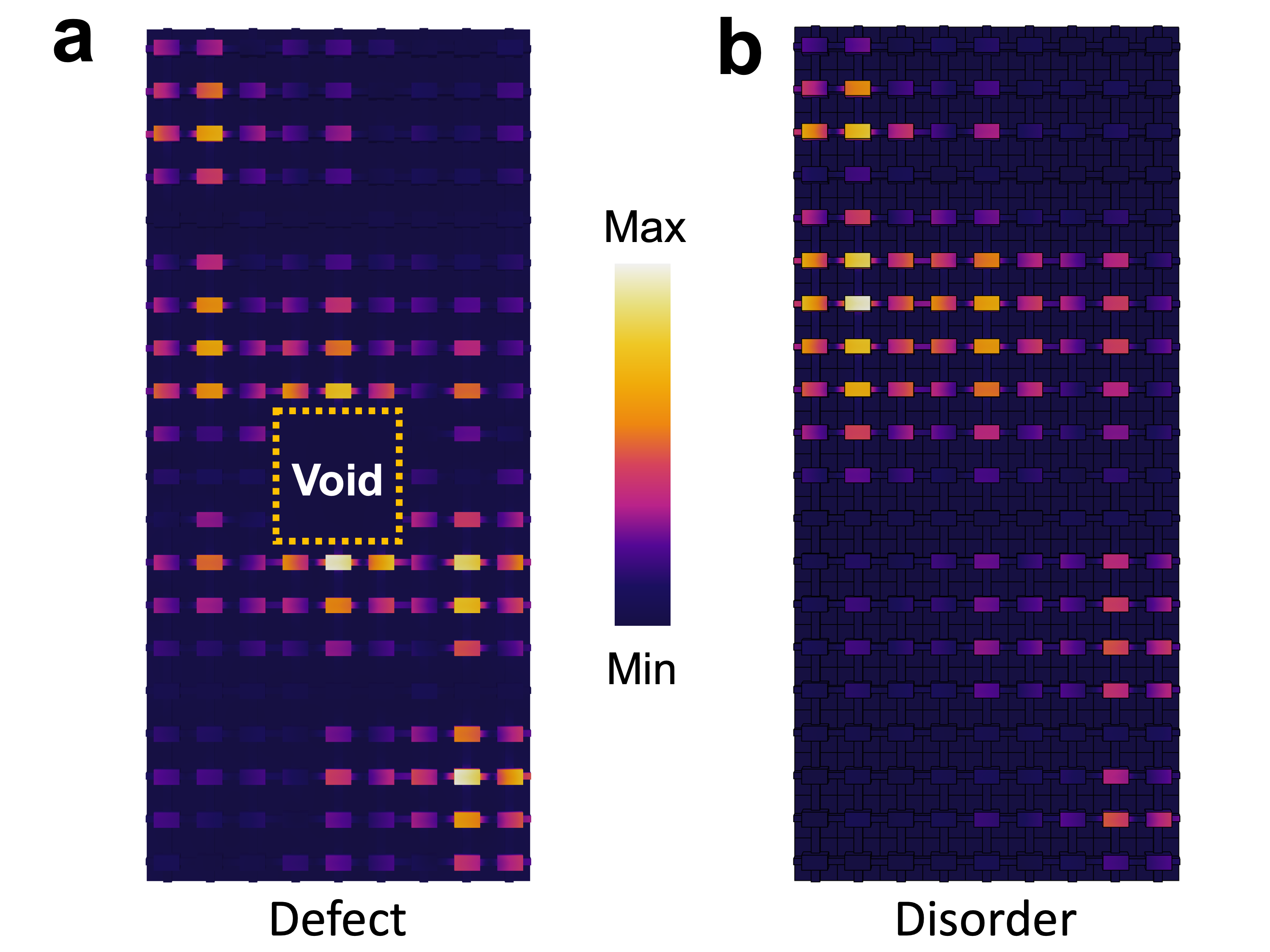}
  \caption{\textbf{Robust topological surface wave pumping}. \textbf{a} Eigenmode of the defective structure where 3 by 3 unit cells of pillars are removed in the center of the structure at 41.75 kHz. The defect is constructed by removing resonators and pillars in the dotted line box \textbf{b} Eigenmode of the disordered structure where a random normal distribution of errors are added on all geometric parameters at 41.88 kHz.}
  \label{defect&disorder}
\end{figure} 

\begin{figure}
  \centering
  \includegraphics[width=\linewidth]{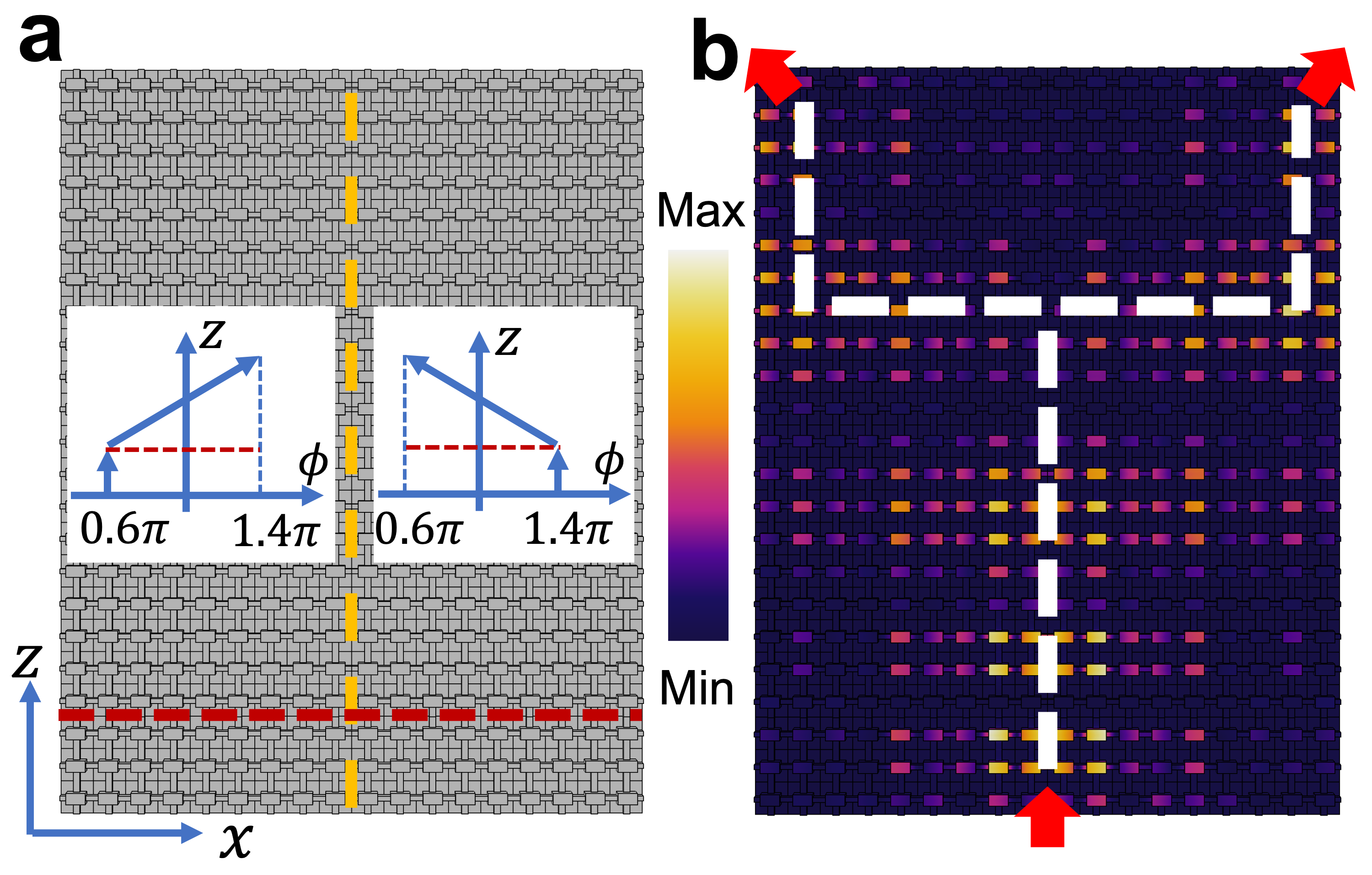}
  \caption{\textbf{Topologically protected surface wave splitter. } \textbf{a} Schematic of surface wave pumping system and the corresponding phase modulation functions. \textbf{b} The displacement field distribution of the surface wave splitter. The surface wave is injected at the center of the bottom edge with the frequency at $f=$41.88~\text{kHz}.}
  \label{splitter}
\end{figure}

We conduct the transient analysis for better showcasing the pumping process. In particular, the right edge mode $\boldsymbol{\varphi}_n(0)$ at the bottom supercell is excited by using a series of piezoelectric patches, each attached on one side of each resonator (see \textcolor{blue}{Methods experimental protocol}). The polarization directions of these piezoelectric patches are identical, while the applied voltages are distributed as $V_0\boldsymbol{\varphi}_n(0) f_z(t)$, where $V_0=1 ~\text{V}$ denotes the voltage amplitude, and $f_z(t)$ is a 50-cycle tone burst signal $f_z(t) = H(50/f_c-t)[1-\cos(2\pi f_c t/50)]\sin(2\pi f_c t)$ (top panel of Fig. \ref{transient}), with $H(t)$ being the Heaviside function and $f_c=41.88$ kHz. Figure \ref{transient}a-c displays the snapshots of surface wave propagation at representative time instants. Initially, the right edge mode is excited on the bottom at $t=0.5 \ \text{ms}$ (Fig. \ref{transient}a). As time progresses, the wave packet propagates in the synthetic dimension $z$ and transitions into the bulk mode at $t=2.5 \ \text{ms}$ (see Fig. \ref{transient}b). Eventually, the left edge mode is well formed on the top of the cuboid at $t=4 \ \text{ms}$ (Fig. \ref{transient}c). The wave packet will follow the same evolution path transitioning from the left edge mode back to the right one if the transient simulation continues. A more detailed demonstration can be found in \textcolor{blue}{Supplementary Movies}.

\noindent \textbf{Robustness of topological surface wave transport.} Indeed, the geometric imperfections in sample fabrication are inevitable due to the errors of millers of CNC machines, as minor discrepancies between simulations and measurement are visible in Fig. \ref{fig2}a, b. Nevertheless, the topological surface wave transport is evidently observed, thanks to some intriguing wave transport characteristics, such as robustness against geometrical impurities or defects. To illustrate this, Fig. \ref{defect&disorder}a shows a lattice defect constructed by removing $3\times 3$ pillars in the middle of the structure. The corresponding eigenfrequency of the EBE mode of the defective cuboid (41.75 kHz) is quite close to that of a perfect cuboid (41.88 kHz). The resulting spatial profile shows that the topological pumping behavior survives and the edge modes can be smoothly pumped from one side to the other despite the large-scale geometric defect. Moreover, we also consider the influence of geometrical disorders. In the sample fabrication, the machining error is about $0.02 \ \text{mm}$ for our sample. Therefore, we introduce errors that satisfy a normal distribution $\mathcal{N}(0 \ \text{mm},0.02 \ \text{mm})$ to the dimensions of all resonators, including their lengths, heights and widths. The spatial profile of the EBE mode with disorders is shown in Fig. \ref{defect&disorder}b. The eigenfrequency of the EBE mode shifts slightly, and the spatial profile agrees with that of the perfect lattice in the $x$-direction, indicating the topological pumping is robust against disorders. However, in the $z$-direction, we see amplitudes of resonators in the top part are larger than amplitudes of resonators in the bottom part, which is actually a sign of Anderson localization. Since along $z$ direction, the displacement field is harmonic, the disorders make the eigenmode localized at the top. And it is observed in the experiment (see Fig. \ref{fig2}a) that the eigenmode is localized at the top. 

\noindent \textbf{Application of topological wave transport as wave splitter.} The surface wave topological pumping is promising for practical applications. To show that, we design a topological split-flow device which performs robust surface wave splitting. As shown in Fig. \ref{splitter}a, the splitter is an assembly of two domains with opposite $\phi$-$z$ distributions, separated by a domain wall (yellow in Fig. \ref{splitter}a). Specifically, the upper section with 20 supercells of the left domain is designed with a linear $\phi$ transition from $0.6\pi$ to $1.4\pi$, whereas that of the right domain is assigned an opposite $\phi$ transition, i.e. from $1.4\pi$ to $0.6\pi$. As for the lower section with 3 supercells, $\phi$ keeps constant at $0.6\pi$ and $1.4\pi$ for the left and right domains, respectively. The excitation is located in the middle of the bottom. Within the lower section of the surface wave splitter, there exists a localized interface mode. As the incidence reaches the upper half, due to the opposite gradients of $\phi$, the interface mode is split into two components, each following the typical EBE evolution but tracing opposite paths. Thanks to topological protection, the propagation is immune against back reflection from  the discontinuity of the upper and lower sections. As such, our design, based on phason engineering and topological pumping, provides an avenue for the application of elastic surface wave beam splitters and smart patterning. In addition, our design covers the short-wavelength range such that we have the opportunity to engineer the dispersion with respect to $k_z$ quasi-momentum. This involves modulations along the vertical direction and opens up a new dimension in the design space for surface wave, which is yet to be explored.

\section*{Discussion}

In conclusion, we have evidenced the topological surface wave transport in modulated phononic crystals through edge-to-edge topological pumpings associated with the 2D quantum Hall effects by the physical rendering of synthetic spaces. These observations imply that the system is characterized by a non-zero Chern number and therefore the topological pumping is immune to bulk scattering and exhibits strong protection against design imperfections. The modulated phononic crystals with synthetic spaces offer a platform and route for efficient surface wave topological mode transport by engineering desired patterns on a phason-torus in the finite structure. The phason space augments the physical space and opens a door to higher-dimensional physics in acoustics and mechanics. Although we focused on the elastic implementation using synthetic spaces, our approach can be generalized to other degrees of freedom, such as additional frequency dimensions can also be harnessed for the frequency modulation. Going forward, it will be important to develop and explore such broader connections, as the idea of topological matter in synthetic dimensions is very general and the extension of this approach to other complex orbits is much awaited. At last, we emphasize that, in order to achieve a reasonable adiabatic regime, the number of chains in our experimental set-ups is appreciable and, whereas this is perfectly fine for the demonstration purposes, it could be an obstacle for practical applications. It will be interesting to explore if this strategy can be deployed for our phononic crystals in order to reduce the number of chains needed for the topological pumping of surface wave.



\section*{Methods}

\noindent \textbf{Sample preparation}

\noindent The experimental sample made of aluminum, having Young’s module $E=69~\text{GPa}$, Poisson's ratio $\nu=0.33$, and density $\rho=2700~\text{kg}/\text{m}^3$, is fabricated using the computer numerical control (CNC) milling machine with a manufacturing precision of $0.02~\text{mm}$. It consists of an array of resonators $(6~\text{mm} \times 3.5~\text{mm} \times 10~\text{mm})$ with the number of 9 along the $x$-direction and 20 along the z direction, which are integrated with a cuboid $(150~\text{mm} \times 50 ~\text{mm} \times 200~\text{mm})$. For convenience, each resonator has an address $(i,j)$. In the x direction, resonators $(i,j)$ and $(i+1,j)$ are connected by height-modulated pillars $(4~\text{mm} \times 1.5 ~\text{mm} \times h_{ij})$ which satisfy the protocol $h_{ij}=h_0[1+\Delta_0 \cos(2\pi i/3+\phi_j)]$, where $h_0=7~\text{mm}, \Delta_0=0.15$ is the average thickness of horizontal channels. Besides, $\phi_j = \phi_s+(\phi_f-\phi_s)j/N$, where $\phi_s=0.6\pi,\phi_f=1.4\pi, N=20$. In the $z$-direction, all resonators are connected by pillars of the same size $(2~\text{mm} \times 6.5 ~\text{mm} \times 3.8~\text{mm})$.

\noindent \textbf{Experimental protocol} 

\noindent In the experiment, the sample is supported by four points to mimic the free boundary condition. A piezoelectric ceramic patch (PZT) is attached to the right side of the cuboid to excite the target eigenmode state. A wide spectrum pseudo-random excitation within the probing ranges from $20-50~\text{kHz}$ is generated by a Tektronix AFG3051C arbitrary waveform generator and amplified by a Krohn-Hite high-voltage power amplifier, which is finally applied across the PZT source. A 1D scanning laser doppler vibrometer (SLDV, Polytech PSV-500) is used to measure the vibration velocity of resonators in the $y$-direction, where high-gain reflective tape is stuck on the surface of each resonator to enhance the reflection of the laser. The velocity signal from the vibrometer is further recorded by the PSV-500 data acquisition. Note that the experiment is repeated and averaged 5 times on each resonator of the system to filter out part of the noise. The normalized amplitude spectrum obtained by applying the Fourier transform to the time-domain signals collected at the resonator $(2,18)$ is shown in Fig. 2c. A series of resonant peaks are observed in the frequency range. By checking the mode shape of each resonance peak in the frequency spectrum, the EBE state corresponding to the frequency at 42.45 kHz is identified. Moreover, a full field measurement at 42.45 kHz is conducted by exciting the system with a 200-cycle sine burst, and the same EBE state is measured.

\noindent \textbf{Numerical simulations} 

\noindent The full-wave finite-element method simulations in this work are all performed using the commercial software COMSOL Multiphysics. The material of 3D structure is implemented by Aluminum [solid] from COMSOL Material Library. Eigenfrequency analysis within the “Solid Mechanics” is carried out to calculate the eigenfrequencies and eigenmode of the unit cell, supercell, and cuboid. The boundary conditions for all the cases are set as free boundary conditions except for Floquet periodicity boundary conditions of the unit cell along $x$ and $z$ direction, and of the supercell along $z$ direction. For the transient analysis in Fig. \ref{transient}, time-dependent analysis in the "Solid Mechanics" are used. Piezoelectric patches (PZT-5H in COMSOL Material Library), are attached on one side of each resonator of the bottom supercell. The polarization directions of these piezoelectric patches are identical, while the applied voltages are distributed as $V \boldsymbol{\varphi}_n(0) f_z (t)$.

\begin{acknowledgments}
This work is supported by the Air Force Office of Scientific Research under Grant No. AF 9550-18-1-0342 and AF 9550-20-1-0279 with Program Manager Dr. Byung-Lip (Les) Lee, the Army Research Office under Grant No. W911NF-18-1-0031 with Program Manager Dr. Daniel P Cole, and the NSF CMMI under Award No. 1930873. Rui Zhu acknowledges support from the National Natural Science Foundation of China (NSFC) under Grant No. 11991033. Emil Prodan acknowledges support from the U.S. National Science Foundation through the grants DMR-1823800 and CMMI-2131760.
\end{acknowledgments}

\section*{Reference}
\bibliographystyle{unsrt}
\bibliography{manu_nonlocal}

\begin{thebibliography}{10}

\bibitem{hasan2010colloquium}
M~Zahid Hasan and Charles~L Kane.
\newblock Colloquium: topological insulators.
\newblock {\em Reviews of modern physics}, 82(4):3045, 2010.

\bibitem{qi2011topological}
Xiao-Liang Qi and Shou-Cheng Zhang.
\newblock Topological insulators and superconductors.
\newblock {\em Reviews of Modern Physics}, 83(4):1057, 2011.

\bibitem{chiu2016classification}
Ching-Kai Chiu, Jeffrey~CY Teo, Andreas~P Schnyder, and Shinsei Ryu.
\newblock Classification of topological quantum matter with symmetries.
\newblock {\em Reviews of Modern Physics}, 88(3):035005, 2016.

\bibitem{rachel2018interacting}
Stephan Rachel.
\newblock Interacting topological insulators: a review.
\newblock {\em Reports on Progress in Physics}, 81(11):116501, 2018.

\bibitem{lindner2011floquet}
Netanel~H Lindner, Gil Refael, and Victor Galitski.
\newblock Floquet topological insulator in semiconductor quantum wells.
\newblock {\em Nature Physics}, 7(6):490--495, 2011.

\bibitem{xu2014observation}
Yang Xu, Ireneusz Miotkowski, Chang Liu, Jifa Tian, Hyoungdo Nam, Nasser
  Alidoust, Jiuning Hu, Chih-Kang Shih, M~Zahid Hasan, and Yong~P Chen.
\newblock Observation of topological surface state quantum hall effect in an
  intrinsic three-dimensional topological insulator.
\newblock {\em Nature Physics}, 10(12):956--963, 2014.

\bibitem{yoshimi2015quantum}
R~Yoshimi, A~Tsukazaki, Y~Kozuka, J~Falson, KS~Takahashi, JG~Checkelsky,
  N~Nagaosa, M~Kawasaki, and Y~Tokura.
\newblock Quantum hall effect on top and bottom surface states of topological
  insulator (bi1- xsbx) 2te3 films.
\newblock {\em Nature communications}, 6(1):1--6, 2015.

\bibitem{haldane2008possible}
FDM Haldane and S~Raghu.
\newblock Possible realization of directional optical waveguides in photonic
  crystals with broken time-reversal symmetry.
\newblock {\em Physical review letters}, 100(1):013904, 2008.

\bibitem{wang2009observation}
Zheng Wang, Yidong Chong, John~D Joannopoulos, and Marin Solja{\v{c}}i{\'c}.
\newblock Observation of unidirectional backscattering-immune topological
  electromagnetic states.
\newblock {\em Nature}, 461(7265):772--775, 2009.

\bibitem{rechtsman2013photonic}
Mikael~C Rechtsman, Julia~M Zeuner, Yonatan Plotnik, Yaakov Lumer, Daniel
  Podolsky, Felix Dreisow, Stefan Nolte, Mordechai Segev, and Alexander
  Szameit.
\newblock Photonic floquet topological insulators.
\newblock {\em Nature}, 496(7444):196--200, 2013.

\bibitem{hafezi2013imaging}
Mohammad Hafezi, S~Mittal, J~Fan, A~Migdall, and JM~Taylor.
\newblock Imaging topological edge states in silicon photonics.
\newblock {\em Nature Photonics}, 7(12):1001--1005, 2013.

\bibitem{lu2014topological}
Ling Lu, John~D Joannopoulos, and Marin Solja{\v{c}}i{\'c}.
\newblock Topological photonics.
\newblock {\em Nature photonics}, 8(11):821--829, 2014.

\bibitem{ozawa2019topological}
Tomoki Ozawa, Hannah~M Price, Alberto Amo, Nathan Goldman, Mohammad Hafezi,
  Ling Lu, Mikael~C Rechtsman, David Schuster, Jonathan Simon, Oded Zilberberg,
  et~al.
\newblock Topological photonics.
\newblock {\em Reviews of Modern Physics}, 91(1):015006, 2019.

\bibitem{fleury2016floquet}
Romain Fleury, Alexander~B Khanikaev, and Andrea Alu.
\newblock Floquet topological insulators for sound.
\newblock {\em Nature communications}, 7(1):1--11, 2016.

\bibitem{yang2015topological}
Zhaoju Yang, Fei Gao, Xihang Shi, Xiao Lin, Zhen Gao, Yidong Chong, and Baile
  Zhang.
\newblock Topological acoustics.
\newblock {\em Physical review letters}, 114(11):114301, 2015.

\bibitem{mousavi2015topologically}
S~Hossein Mousavi, Alexander~B Khanikaev, and Zheng Wang.
\newblock Topologically protected elastic waves in phononic metamaterials.
\newblock {\em Nature communications}, 6(1):1--7, 2015.

\bibitem{souslov2017topological}
Anton Souslov, Benjamin~C Van~Zuiden, Denis Bartolo, and Vincenzo Vitelli.
\newblock Topological sound in active-liquid metamaterials.
\newblock {\em Nature Physics}, 13(11):1091--1094, 2017.

\bibitem{wang2015topological}
Pai Wang, Ling Lu, and Katia Bertoldi.
\newblock Topological phononic crystals with one-way elastic edge waves.
\newblock {\em Physical review letters}, 115(10):104302, 2015.

\bibitem{nash2015topological}
Lisa~M Nash, Dustin Kleckner, Alismari Read, Vincenzo Vitelli, Ari~M Turner,
  and William~TM Irvine.
\newblock Topological mechanics of gyroscopic metamaterials.
\newblock {\em Proceedings of the National Academy of Sciences},
  112(47):14495--14500, 2015.

\bibitem{nassar2018quantization}
H~Nassar, H~Chen, AN~Norris, and GL~Huang.
\newblock Quantization of band tilting in modulated phononic crystals.
\newblock {\em Physical Review B}, 97(1):014305, 2018.

\bibitem{chen2019mechanical}
H~Chen, LY~Yao, H~Nassar, and GL~Huang.
\newblock Mechanical quantum hall effect in time-modulated elastic materials.
\newblock {\em Physical Review Applied}, 11(4):044029, 2019.

\bibitem{Rosa2019}
Matheus~I.N. Rosa, Raj~Kumar Pal, Jos{\'{e}}~R.F. Arruda, and Massimo Ruzzene.
\newblock {Edge States and Topological Pumping in Spatially Modulated Elastic
  Lattices}.
\newblock {\em Physical Review Letters}, 123(3):1--6, 2019.

\bibitem{Riva2020}
Emanuele Riva, Matheus~I.N. Rosa, and Massimo Ruzzene.
\newblock {Edge states and topological pumping in stiffness-modulated elastic
  plates}.
\newblock {\em Physical Review B}, 101(9), 2020.

\bibitem{grinberg2020robust}
Inbar~Hotzen Grinberg, Mao Lin, Cameron Harris, Wladimir~A Benalcazar,
  Christopher~W Peterson, Taylor~L Hughes, and Gaurav Bahl.
\newblock Robust temporal pumping in a magneto-mechanical topological
  insulator.
\newblock {\em Nature communications}, 11(1):1--9, 2020.

\bibitem{price2015four}
Hannah~M Price, Oded Zilberberg, Tomoki Ozawa, Iacopo Carusotto, and Nathan
  Goldman.
\newblock Four-dimensional quantum hall effect with ultracold atoms.
\newblock {\em Physical review letters}, 115(19):195303, 2015.

\bibitem{lohse2016thouless}
Michael Lohse, Christian Schweizer, Oded Zilberberg, Monika Aidelsburger, and
  Immanuel Bloch.
\newblock A thouless quantum pump with ultracold bosonic atoms in an optical
  superlattice.
\newblock {\em Nature Physics}, 12(4):350--354, 2016.

\bibitem{lohse2018exploring}
Michael Lohse, Christian Schweizer, Hannah~M Price, Oded Zilberberg, and
  Immanuel Bloch.
\newblock Exploring 4d quantum hall physics with a 2d topological charge pump.
\newblock {\em Nature}, 553(7686):55--58, 2018.

\bibitem{Kraus2012}
Yaacov~E. Kraus, Yoav Lahini, Zohar Ringel, Mor Verbin, and Oded Zilberberg.
\newblock {Topological states and adiabatic pumping in quasicrystals}.
\newblock {\em Physical Review Letters}, 109(10), 2012.

\bibitem{Kraus2013}
Yaacov~E. Kraus, Zohar Ringel, and Oded Zilberberg.
\newblock {Four-dimensional quantum hall effect in a two-dimensional
  quasicrystal}.
\newblock {\em Physical Review Letters}, 111(22):1--5, 2013.

\bibitem{verbin2015topological}
Mor Verbin, Oded Zilberberg, Yoav Lahini, Yaacov~E Kraus, and Yaron Silberberg.
\newblock Topological pumping over a photonic fibonacci quasicrystal.
\newblock {\em Physical Review B}, 91(6):064201, 2015.

\bibitem{zilberberg2018photonic}
Oded Zilberberg, Sheng Huang, Jonathan Guglielmon, Mohan Wang, Kevin~P Chen,
  Yaacov~E Kraus, and Mikael~C Rechtsman.
\newblock Photonic topological boundary pumping as a probe of 4d quantum hall
  physics.
\newblock {\em Nature}, 553(7686):59--62, 2018.

\bibitem{chen2021creating}
Hui Chen, Hongkuan Zhang, Qian Wu, Yu~Huang, Huy Nguyen, Emil Prodan, Xiaoming
  Zhou, and Guoliang Huang.
\newblock Creating synthetic spaces for higher-order topological sound
  transport.
\newblock {\em Nature communications}, 12(1):1--10, 2021.

\bibitem{Long2019}
Yang Long and Jie Ren.
\newblock {Floquet topological acoustic resonators and acoustic Thouless
  pumping}.
\newblock {\em The Journal of the Acoustical Society of America},
  146(1):742--747, 2019.

\bibitem{chen2021landau}
Ze-Guo Chen, Weiyuan Tang, Ruo-Yang Zhang, Zhaoxian Chen, and Guancong Ma.
\newblock Landau-zener transition in the dynamic transfer of acoustic
  topological states.
\newblock {\em Physical Review Letters}, 126(5):054301, 2021.

\bibitem{rayleigh1885waves}
Lord Rayleigh.
\newblock On waves propagated along the plane surface of an elastic solid.
\newblock {\em Proceedings of the London mathematical Society}, 1(1):4--11,
  1885.

\bibitem{campbell2012surface}
Colin Campbell.
\newblock {\em Surface acoustic wave devices and their signal processing
  applications}.
\newblock Elsevier, 2012.

\bibitem{morgan2010surface}
David Morgan.
\newblock {\em Surface acoustic wave filters: With applications to electronic
  communications and signal processing}.
\newblock Academic Press, 2010.

\bibitem{hashimoto2000surface}
Ken-ya Hashimoto and Ken-Ya Hashimoto.
\newblock {\em Surface acoustic wave devices in telecommunications}, volume
  116.
\newblock Springer, 2000.

\bibitem{jin2017pillar}
Yabin Jin, Bernard Bonello, Rayisa~P Moiseyenko, Yan Pennec, Olga Boyko, and
  Bahram Djafari-Rouhani.
\newblock Pillar-type acoustic metasurface.
\newblock {\em Physical Review B}, 96(10):104311, 2017.

\bibitem{packo2019inverse}
Pawel Packo, Andrew~N Norris, and Daniel Torrent.
\newblock Inverse grating problem: Efficient design of anomalous flexural wave
  reflectors and refractors.
\newblock {\em Physical Review Applied}, 11(1):014023, 2019.

\bibitem{wu2021non}
Qian Wu, Hui Chen, Hussein Nassar, and Guoliang Huang.
\newblock Non-reciprocal rayleigh wave propagation in space--time modulated
  surface.
\newblock {\em Journal of the Mechanics and Physics of Solids}, 146:104196,
  2021.

\bibitem{wang2022extended}
Ji-Qian Wang, Zi-Dong Zhang, Si-Yuan Yu, Hao Ge, Kang-Fu Liu, Tao Wu, Xiao-Chen
  Sun, Le~Liu, Hua-Yang Chen, Cheng He, et~al.
\newblock Extended topological valley-locked surface acoustic waves.
\newblock {\em Nature communications}, 13(1):1--8, 2022.

\bibitem{zhang2021topological}
Zi-Dong Zhang, Si-Yuan Yu, Hao Ge, Ji-Qian Wang, Hong-Fei Wang, Kang-Fu Liu,
  Tao Wu, Cheng He, Ming-Hui Lu, and Yan-Feng Chen.
\newblock Topological surface acoustic waves.
\newblock {\em Physical Review Applied}, 16(4):044008, 2021.

\bibitem{ChengPRL2020}
W.~Cheng, E.~Prodan, and C.~Prodan.
\newblock Experimental demonstration of dynamic topological pumping across
  incommensurate acoustic meta-crystals.
\newblock {\em Phys. Rev. Lett.}, 125:224301, 2020.

\bibitem{holmes2012introduction}
M.H. Holmes.
\newblock {\em Introduction to Perturbation Methods}.
\newblock Texts in Applied Mathematics. Springer New York, 2012.

\bibitem{NiCommPhys2019}
Xiang Ni, Kai Chen, Matthew Weiner, David~J. Apigo, Camelia Prodan, Andrea
  Alù, Emil Prodan, and Alexander~B. Khanikaev.
\newblock Observation of hofstadter butterfly and topological edge states in
  reconfigurable quasi-periodic acoustic crystals.
\newblock {\em Communications Physics}, 2:55, 2019.

\end{thebibliography}

\end{document}